# Induced-charge Electrophoresis of Metallo-dielectric Particles

Sumit Gangwal,[1] Olivier J. Cayre,[1] Martin Z. Bazant,[2] and Orlin D. Velev[1,*]

[1]*Department of Chemical and Biomolecular Engineering, North Carolina State University, Raleigh, North Carolina, 27695, USA*

[2]*Department of Mathematics and Institute for Soldier Nanotechnologies, Massachusetts Institute of Technology, Cambridge, Massachusetts, 02139, USA*

The application of AC electric fields in aqueous suspensions of anisotropic particles leads to unbalanced liquid flows and nonlinear, induced-charge electrophoretic (ICEP) motion. We report experimental observations of the motion of "Janus" microparticles with one dielectric and one metal-coated hemisphere induced by uniform fields of frequency 100 Hz - 10 kHz in NaCl solutions. The motion is perpendicular to the field axis and persists after particles are attracted to a glass wall. This phenomenon may find applications in microactuators, microsensors, and microfluidic devices.

Nonlinear electrokinetic phenomena are widely used to manipulate colloids and drive flows in microfluidic devices. The liquid and particle velocities typically depend on the strength of the applied field squared and are commonly driven by alternating current (AC) to avoid Faradaic reactions. The classical example is dielectrophoresis (DEP), where a net electrostatic force causes particle motion in a non-uniform AC field. Polarization of the ionic double layers can also lead to nonlinear electro-osmotic flows at low frequencies (kHz), as first described by Murtsovkin and coworkers [1,2]. Bazant and Squires [3,4] conceptually unified this phenomenon with AC electro-osmosis at electrodes, first described by Ramos *et al.* [5] and Ajdari [6], and suggested the term "induced-charge electro-osmosis" (ICEO) to describe all flows resulting from the action of an applied electric field on its own induced diffuse charge near a polarizable surface. They also predicted how broken symmetries could cause polarizable particles to move in electric fields by "induced-charge electrophoresis" (ICEP) [3,7].

Although ICEO flows have been observed in diverse settings [1,5,8,9], ICEP motion of colloids is largely unexplored. In what may be the only prior experimental work, Murtsovkin and Mantrov [10] observed that the motion of quartz particles of irregular shapes varies with the square of the applied field, but did not provide a theory. General mobility relations for homogenous non-spherical conducting particles have been derived [11] and ICEP motion has been calculated for arbitrary shape perturbations [7], as well as rod-like spheroidal shapes [12].

The synthesis of "Janus" particles (whose halves are physically or chemically different) and their application in novel materials is a relatively new but rapidly expanding research field [13-17]. Anisotropic particles with two hemispheres of different polarizabily or conductance have been produced by thermal evaporation [18] or gold sputtering [19]. The mobility of such particles in external fields, however, has not been investigated systematically.

In this Letter, we report how metallo-dielectric Janus particles suspended in water perform ICEP motion in directions perpendicular to a uniform AC field. This unusual phenomenon, which cannot be attributed to DEP, is the first experimental observation of an effect predicted theoretically by Squires and Bazant [7], although we find important differences as well. Using the standard low-voltage model for thin double layers, Squires and Bazant predicted that a metallic sphere with a hemispherical dielectric coating oriented perpendicular to the field will move by ICEP in the direction of its dielectric end at a velocity,

$$U_{ICEP} = \frac{9}{64} \frac{\varepsilon R E_0^2}{\eta(1+\delta)} \qquad (1)$$

where $\varepsilon$ is the permittivity and $\eta$ the viscosity of the bulk solvent, $R$ is the radius of the sphere, $E_0$ is the field amplitude, and $\delta$ is the ratio of the differential capacitances of the compact and diffuse layers. Eq. 1 also describes the time-averaged velocity in a square-wave AC field at low frequency, as in our experiments.

The Janus particles studied were prepared by partially coating dielectric microspheres with a conductive layer of gold. Aqueous suspensions of surfactant-free sulfate-stabilized polystyrene latex spheres (diameter, $D$ = 4.0, 5.7, and 8.7 μm) were purchased from Interfacial Dynamics Corporation (OR). After centrifuging and washing with ultra-pure Milli-Q water the particles were concentrated and deposited in monolayers on pre-cleaned glass slides by our convective assembly method [20]. The top sides of the dried particle monolayers were coated with 10 nm of chromium followed by 20 nm of gold in a metal evaporator. The "Janus" particles formed [Fig. 1(a)] were then redispersed in Milli-Q water by mild sonication.

Diluted particle suspensions (approximately ~ 0.01 % solids) were injected in the thin chamber of an experimental cell facing two electrodes [Fig. 1(b)]. The field-driven motion of the particles in the chamber was recorded with a digital camera from above using an Olympus BX-61 optical microscope. The average particle velocities were calculated for 15-25 particles for each experimental condition. The standard error of each average is reported in all plots.

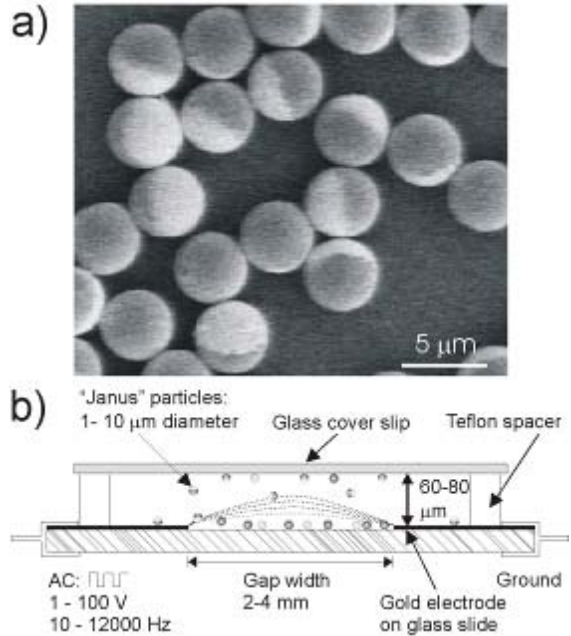

FIG. 1. (a) SEM image of $4.0\,\mu m$ polystyrene particles partially coated with gold. The gold-coated hemispheres appear brighter due to their higher conductance. (b) Schematic of the experimental set-up (not drawn to scale). Two gold electrodes are deposited on the bottom plate. A Teflon spacer sustains a $60\text{-}80\,\mu m$ gap between the bottom and the top microscope cover slip. The particle suspension is surrounded by a hydrophobic ring and confined in this thin chamber by capillarity.

The Janus particles were subjected to AC fields between 100 - 320 V/cm at a frequency between 0.1 - 10 kHz. When the electric field in the cell was turned on, the particles oriented such that the plane between their hemispheres (gold-coated/conductive hemisphere appearing dark and bare/dielectric hemisphere appearing light) aligned in the direction of the electric field. The rotation of the particles to this stable orientation can be caused at least partly by DEP, since it results in the largest induced dipole moment, aligned with the electric field direction. ICEO flows may also contribute, however, since the associated hydrodynamic torque is predicted to rotate the particle to the same orientation as DEP [7]. The particles then moved normal to the applied electric field with their polystyrene hemisphere forward [Fig. 2(a)]. This motion cannot be attributed to DEP. The larger particles moved at a higher velocity than the smaller ones [Fig. 2(a)]. The phenomenon is nicely illustrated in the movies provided as supplementary material.

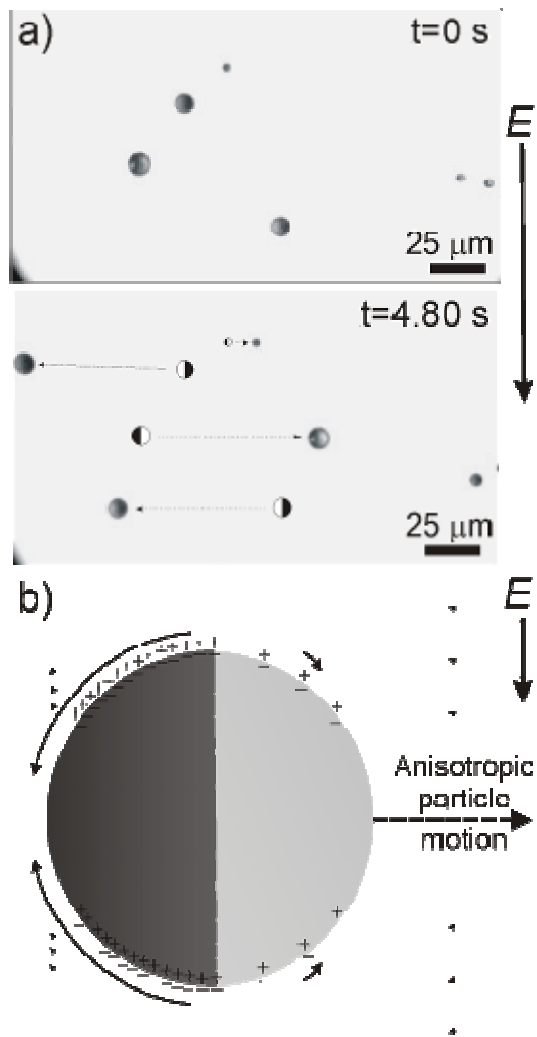

FIG. 2. (a) Optical micrographs from different frames of a recording of the position and orientation of Janus particles of three different diameters (4.0, 5.7 and 8.7 $\mu$m) in an AC field of amplitude 140 V/cm and frequency 1 kHz. The two particles on the right side in the top image have moved out of the field of view and another particle has moved into view in the bottom image approximately 5 s later. (b) Schematic of a particle in one-half cycle of AC electric field in the stable configuration. The electric double layer on the gold side (black hemisphere) is more strongly polarized and thus drives a stronger ICEO slip (arrows) than the polystyrene side, resulting in ICEP motion in the direction of the dielectric side.

These observations of a new mode of particle mobility are qualitatively consistent with ICEP [7]. At low AC frequency, there is ample time for double-layer charging to screen the electric field, leading to the induced charge distribution sketched in Fig. 2(b). In each half cycle of the AC field, all of the induced charges change sign in phase with the field, thus yielding the same ICEO flow, varying as $E^2$. For a homogeneous particle, the flow profile is quadrupolar, drawing fluid in along the field axis and ejecting it radially from the equator [3]. For our Janus particles in the stable orientation, however, the induced charge, and hence the ICEO flow, are much stronger around the metal-coated hemisphere, driving the particle to move by ICEP in the direction of the dielectric hemisphere, normal to the electric field.

Surprisingly, the moving Janus spheres appear to be dynamically attracted to the glass walls. While the particles performed transverse motion they remained in stable trajectories, apparently within a few particle diameters of the top or bottom walls. The experimentally observed behavior is likely due to the asymmetry of the Janus particles, since homogenous particles are expected to push away from insulating walls due to ICEO flows [21]. The near-wall trajectories allowed convenient experimental measurements of the velocity of particles moving along the top wall of the cell. Since Eq. (1) was derived for a particle in an infinite fluid, we mainly test the velocity scaling with $E_0$, $c_0$, and $R$. We were not able to measure precisely the velocity of particles moving freely in the liquid (before encountering either wall) because of experimental difficulties with keeping the particles in focus and estimating the vertical component of their velocity.

The data for the dependence of the particle velocity on the field strength and salt concentration are plotted in Fig. 3. In ultra-pure water and in 0.1 mM NaCl, the particle velocity scales as $U_{ICEP} = k\left(\frac{9\varepsilon R}{64\eta}\right)E_0^2$, in accordance with eq. 1. The fitted coefficient $k = 0.090 \pm 0.002$ for $c_0 = 0.1$ mM implies $\delta \approx 10$. This value is somewhat larger than in prior work on ICEO flow in KCl [9], but the fit must be corrected for wall interactions, which reduce the ICEP velocity (see below). With increasing NaCl concentration, the field scaling displays an apparent offset for a critical field required to drive the motion, which is not predicted by the theory. More importantly, the ICEP velocity decreases rapidly as the electrolyte concentration is gradually increased from 0.1 mM to 3 mM, extrapolating to zero velocity at approx. 10 mM. The weak concentration dependence in (1) through $\delta \propto \sqrt{c_0}$ (via the Debye length) cannot explain this dramatic flow suppression.

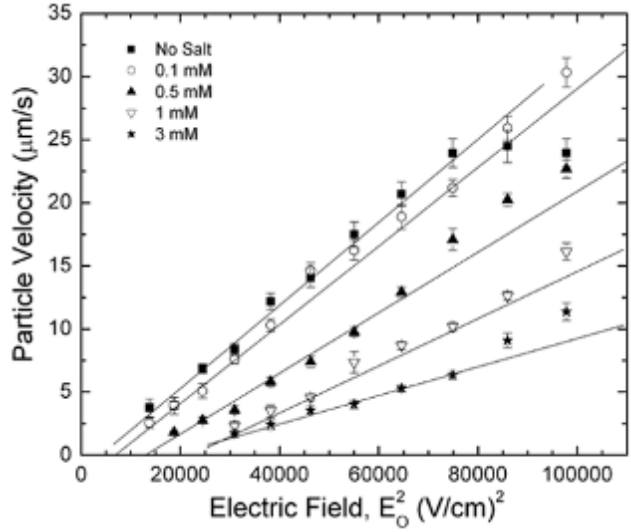

FIG. 3. Velocity of 5.7 μm Janus particles as a function of electric field intensity squared ($E_0^2$) for various NaCl concentrations at 1 kHz. The linear fits generally agree with the experimental values for low intensities and start deviating for $E_0 > 280$ V/cm.

The existing ICEO theory can only be justified for dilute solutions at "small" applied voltages [4], where the induced potential drop across the diffuse layer is less than the thermal voltage $kT/e = 25$ mV. It assumes that the velocity will follow Smoluchowski's formula for electro-osmotic slip mobility,

$$u_s = -\frac{\varepsilon \zeta(E) E}{\eta} \qquad (2)$$

where $\zeta(E)$ is the (induced) potential drop $\psi$ at the no-slip plane, relative to the neutral bulk solution just outside the double layer. For the 5.7 μm particles, the background voltage $V_b = ER$ across each hemisphere varies from 1.33 to 3.55 $kT/e$, so our experiments probe the transition to "large" applied voltages. The functional dependence of such ICEO flows on the $\zeta$-potential has been recently expressed by Bazant *et al.* considering steric effects of counterion crowding [22]. The ICEP velocity of the 5.7 μm Janus particles decayed at NaCl concentrations above 0.1 mM (Fig. 3). The critical concentration where the velocity extrapolates to zero is remarkably small, $c_{max} \approx 10$ mM. Similar concentration dependence has been observed in microfluidic pumping by AC electro-osmosis [23], but we are not aware of any prior observation in colloids. It might be a universal feature of ICEO flows [22].

We also investigated the dependence of the ICEP velocity on the AC field frequency. ICEO flows around polarizable objects persist only in a certain band of driving frequencies, $\tau_e^{-1} \leq \omega \leq \tau_p^{-1}$ [4]. The upper limit is set by the characteristic "RC time" $\tau_p = \frac{\lambda R}{D}$ for the formation of the induced screening cloud on the particle, where $\lambda$ is the Debye length and $D$ is the ion diffusion coefficient. The lower limit is set by the analogous charging time of the electrodes, $\tau_e = \frac{\lambda L}{D}$, where $2L$ is the electrode separation. The estimated upper and lower characteristic frequencies for this experiment are $\tau_e^{-1} \approx 20$ Hz and $\tau_p^{-1} \approx 12$ kHz. The particle velocities measured indeed decreased when approaching both characteristic frequencies (Supplementary Fig. 1). Finally, we characterized the dependence of the ICEP velocity on particle size. At $E_0 = 300$ V/cm, $c_0 = 0.1$ mM and $\omega \tau_p \ll 1$, the average velocity increases roughly linearly with particle diameter at small sizes but seems to reach a constant value around 8 μm (Fig. 4), which corresponds to an induced voltage $E_0 R = 0.24$ V $\approx 10 kT/e$, well into the nonlinear regime. The linear fit from Eq. (1) yields $k = 0.073 \pm 0.002$ or $\delta \approx 12$, which is consistent with the fit of $\delta \approx 10$ obtained from the $E_0^2$ scaling in Fig. 3.

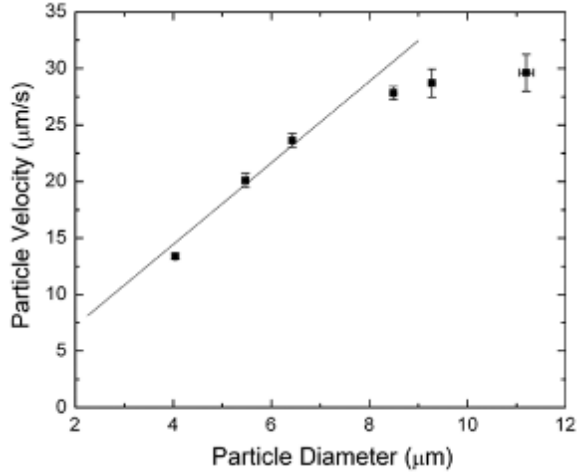

FIG. 4. Average velocity of different size particles as a function of their average diameter at 300 V/cm in 0.1 mM NaCl at a field frequency of 1 kHz. The data points were obtained by determining the velocity of 286 particles grouped by diameters within a 1.5 μm range. The line is a fit to Eq. (1) described in the main text.

The strong ICEP effect observed here is in semi-quantitative agreement with the theory. The results for velocity for low electrolyte concentrations are consistent with the scalings of Eq. 1, though smaller by up to an order of magnitude $k^{-1} = 11 - 13$ (for $\delta = 0$). The smaller velocities are likely a result of wall-particle interactions. The Janus particles motion in stable near-wall trajectories may be due to tilting of the dielectric hemisphere to face the wall, caused by asymmetric ICEO flow [24]. If double-layer repulsion between latex and glass is present, the model yielding Eq. 1 can predict ICEP translation at tilt angles up to 45° and at velocities ≈ 30% smaller than in the bulk, in which case the experimental data can be fit with δ = 7. Indications of tilting can be seen in the experiment, but detecting and measuring small tilt angles from the images is not possible. Other effects may be present as well. Complex ICEO (electrohydrodynamic) flows on electrodes are known to attract particles and drive their self-assembly [25,26], but their importance in this different geometry is yet to be evaluated.

In summary, we demonstrated that metallo-dielectric Janus particles move perpendicular to a uniform AC electric field as a result of induced-charge electrophoretic force. This is, to our knowledge, the first report of this fascinating phenomenon that can now guide further development of the theory. The propelling metallo-dielectric particles could eventually be used as microscopic mixers, "shuttles", self-propelling on-chip sensors, microactuators and microsensors in MEMS devices, optoelectronic devices or for separations in microfluidic devices.

This research was supported by an NSF-NIRT grant and Camille and Henry Dreyfus Teacher-Scholar award (NCSU) and by the U.S. Army through the Institute for Soldier Nanotechnologies (ARO contract No. DAAD-19-02-0002, MIT).

* Electronic address: odvelev@unity.ncsu.edu

# Induced-charge electrophoresis of metallo-dielectric particles

Sumit Gangwal, Olivier J. Cayre, Martin Z. Bazant, and Orlin D. Velev[*]

* Electronic address: odvelev@unity.ncsu.edu

## *Description of the Supplementary Movies*

**ICEP_Motion-3Sizes.wmv**

Janus particles of diameter 4.0, 5.7, and 8.7 $\mu m$ moving with dielectric hemisphere forward by ICEP in the direction normal to AC electric field of 300 V/cm and 1 kHz frequency at the top of the experimental cell (of 60 $\mu m$ thickness). The 8.7 $\mu m$ particle moves faster than the 5.7 $\mu m$ particles, which in turn move at a higher velocity than the 4.0 $\mu m$ particles.

**ICEP_Motion-DipoleAggregation.wmv**

Three gold-coated polystyrene particles (5.7 $\mu m$ in diameter) moving by ICEP in an AC field of 125 V/cm and 5 kHz frequency at the top of the experimental cell (60 $\mu m$ thickness). Two of the particles come together, the induced dipoles align/attract and the particles sink toward the bottom of the cell due to dielectrophoresis.

**ICEP_Motion-HighConc.wmv**

Janus particles of diameter 5.7 $\mu m$ in an AC field of 300 V/cm and 1 kHz frequency at the top of the experimental cell. Particles move by ICEP in either direction normal to the applied electric field. The higher concentration of particles results in increased particle-particle encounters and interactions.

## *Supplementary Figure*

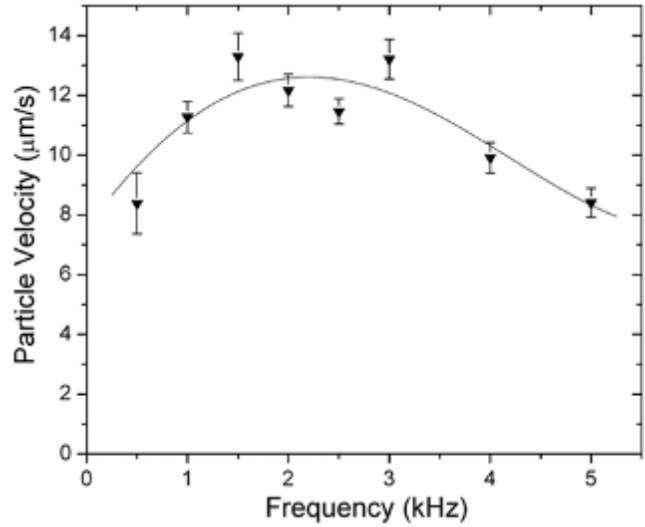

SUPPLEMENTARY FIG. 1. Velocity of 5.7 μm Janus particles as a function of frequency at 200 V/cm in 0.1 mM NaCl. The experimental particle velocities were highest at intermediate frequency and decreased at low and high frequencies. The estimated upper and lower characteristic frequencies for electrode and particle charging in the conditions of this experiment are $\tau_e^{-1} \approx 20$ Hz and $\tau_p^{-1} \approx 12$ kHz, respectively.